\documentclass[aps,prb,numerical,showpacs,reprint,amsfonts,amssymb,amsmath,twocolumms,groupedaddress]{revtex4-1}
\usepackage{graphicx}
\usepackage{multirow}
\usepackage{rotating}
\usepackage{color}

\usepackage[colorlinks=true, linkcolor=blue,citecolor=blue,urlcolor=blue]{hyperref}

\begin{document}
\draft \narrowtext \sloppy
\title{Electronic bands of III-V semiconductor polytypes and their alignment}

\author{Abderrezak Belabbes} \email{abderrezak.belabbes@uni-jena.de}
\author{ Christian Panse}
\author{ J\"urgen Furthm\"uller}
\author{ Friedhelm Bechstedt}
\affiliation{$^1$ Institut f\"{u}r Festk\"{o}rpertheorie und -optik,
Friedrich-Schiller-Universit\"{a}t, Max-Wien-Platz 1, 07743 Jena,
Germany}

\date{\today}
\begin{abstract}
The quasiparticle band structures of four polytypes 3C, 6H, 4H, and 2H of GaP, GaAs, GaSb, InP, InAs, and InSb are computed with high accuracy including spin-orbit interaction applying a recently developed approximate calculation scheme, the LDA-1/2 method. The results are used to derive band offsets $\Delta E_c$ and $\Delta E_v$ for the conduction and valence bands between two polytypes. The alignment of the band structures is based on the branch-point energy $E_{\rm BP}$ for each polytype. The aligned electronic structures are used to explain properties of heterocrystalline but homomaterial junctions. The gaps and offsets allow to discuss spectroscopic results obtained recently for such junctions in III-V nanowires.
\end{abstract}
\maketitle

\section{Introduction}

Functional nanotechnology has become a central task in recent research and technological development. It includes advances in the synthesis of novel nanomaterials. For instance, nanowires (NWs) have attracted much interest due to their potential applications as optically active devices \cite{Duan.Huang.ea:2003:N} and as building blocks for nanocircuits \cite{Nilsson.Thelander.ea:2006:APL,Yang.Yan.ea:2010:NL}. This holds especially for nanowires of III-V compounds which grow in cubic [111] direction. Apart from the nitrides which grow in wurtzite ($wz$) geometry, the most conventional III-V materials such as Ga and In phosphides, arsenides or antimonides crystallize in cubic zinc-blende ($zb$) structure under ambient conditions. However, frequently the [111]-oriented nanowires of conventional III-V compounds exhibit a random intermixing of $zb$ and $wz$ stackings \cite{Joyce.Wong-Leung.ea:2010:NL}.

Controlling the crystallographic phase purity of III-V nanowires is notoriously difficult. However, recently enormous progress has been made in controlled growing of twin-plane or even polytypic superlattices in these III-V nanowires \cite{Algra.Verheijen.ea:2008:N,Caroff.Dick.ea:2009:NN,Dick.Thelander.ea:2010:NL}. Even pure wurtzite nanowires can be grown \cite{Shtrikman.Popovitz-Biro.ea:2009:NL}. Also  the formation of $wz$-GaAs was demonstrated in polycrystalline 
powder samples using pressure treatment \cite{McMahon.Nelmes:2005:PRL}.

Meanwhile, one already speaks about polytypism \cite{Kackell.Wenzien.ea:1994:PRB,Kackell.Wenzien.ea:1994:PRBa} in III-V nanowires \cite{Caroff.Dick.ea:2009:NN,Caroff.Bolinsson.ea:2011:STQE,Kriegner.Panse.ea:2011:NL}. Besides the $zb$ (3C) and $wz$ (2H) crystal structures \cite{Kackell.Wenzien.ea:1994:PRB,Kackell.Wenzien.ea:1994:PRBa} also the 4H or even 6H polytype has been observed for III-V nanorods \cite{Kriegner.Panse.ea:2011:NL,Mandl.Dick.ea:2011:N,Dheeraj.Patriarche.ea:2008:NL,Soshnikov.Cirlin.ea:2008:TPL,Mariager.Sorensen.ea:2007:APL}. The hexagonal polytypes 2H, 4H, and 6H lead to a drastic change of the bonding topology along the cubic [111] or hexagonal [0001] axis \cite{Kriegner.Panse.ea:2011:NL,Panse.Kriegner.ea:2011:PRB} but also to significant changes of the electronic structure, e.g. the fundamental energy gap, with respect to the cubic 3C polytype \cite{Kackell.Wenzien.ea:1994:PRB}.
This especially holds for the transition region between two polytypes, e.g. 3C-$p$H. It can be considered as a homomaterial but heterocrystalline junction \cite{Bechstedt.Kackell:1995:PRL}, which is characterized by band offsets $\Delta E_c$ ($\Delta E_v$) in the conduction (valence) bands similar to a heteromaterial junction. The offsets may form energy barriers for electrons and/or holes. Indeed, indications for such gap variations and band offsets in homomaterial III-V nanowires have been observed in several optical spectroscopies \cite{Caroff.Bolinsson.ea:2011:STQE,Pemasiri.Montazeri.ea:2009:NL,Spirkoska.Arbiol.ea:2009:PRB,Akopian.Patriarche.ea:2010:NL,Ketterer.Heiss.ea:2011:PRB}.

The discovery of the 2H and 4H polytypes in nanowires of conventional III-V compounds in addition to the 3C equilibrium structure asks for the understanding of variation of the electronic structure with the hexagonal bond stacking and the alignment of the band edges between two polytypes of one-and-the-same compound. The trials in the last years toward this understanding by means of almost first-principles calculations were basically restricted to the heterocrystalline junction 3C-2H and the density-functional theory (DFT) which however significantly underestimates the fundamental gap \cite{Murayama.Nakayama:1994:PRB,Yeh.Wei.ea:1994:PRB,Akiyama.Yamashita.ea:2010:NL}. Also the empirical pseudopotential method which, however, cannot yield to band offsets has recently been applied \cite{De.Pryor:2010:PRB}. Improved DFT calculations have been performed for 3C- and 2H-GaAs using a hybrid functional to describe exchange and correlation \cite{Heiss.Conesa-Boj.ea:2011:PRB}. First quasiparticle computations are now available for GaAs and InAs \cite{Zanolli.Fuchs.ea:2007,Cheiwchanchamnangij.Lambrecht:2011:PRB}.

However, systematic quasiparticle studies beyond the density-functional theory for gaps and band discontinuities along the row 3C, 6H, 4H, and 2H with increasing hexagonality of the bonding geometry are missing. Their first-principles calculation is the goal of the present paper. In Sec. II the methods to describe quasiparticle band structures including spin-orbit interaction and to align them by means of the branch-point energy are described. The electronic-structure results are presented and discussed in Sec.~III for Ga and In phosphides, arsenides, and antimonides. The resulting band offsets for the heterocrystalline junctions between two polytypes of the same III-V compound are given in Sec.~IV. Finally, in Sec.~V we give a brief summary and conclusions.

\section{Methodology}
\subsection{Geometries}

\begin{figure}[h!]
\includegraphics[width=0.31\textwidth, angle=-90]{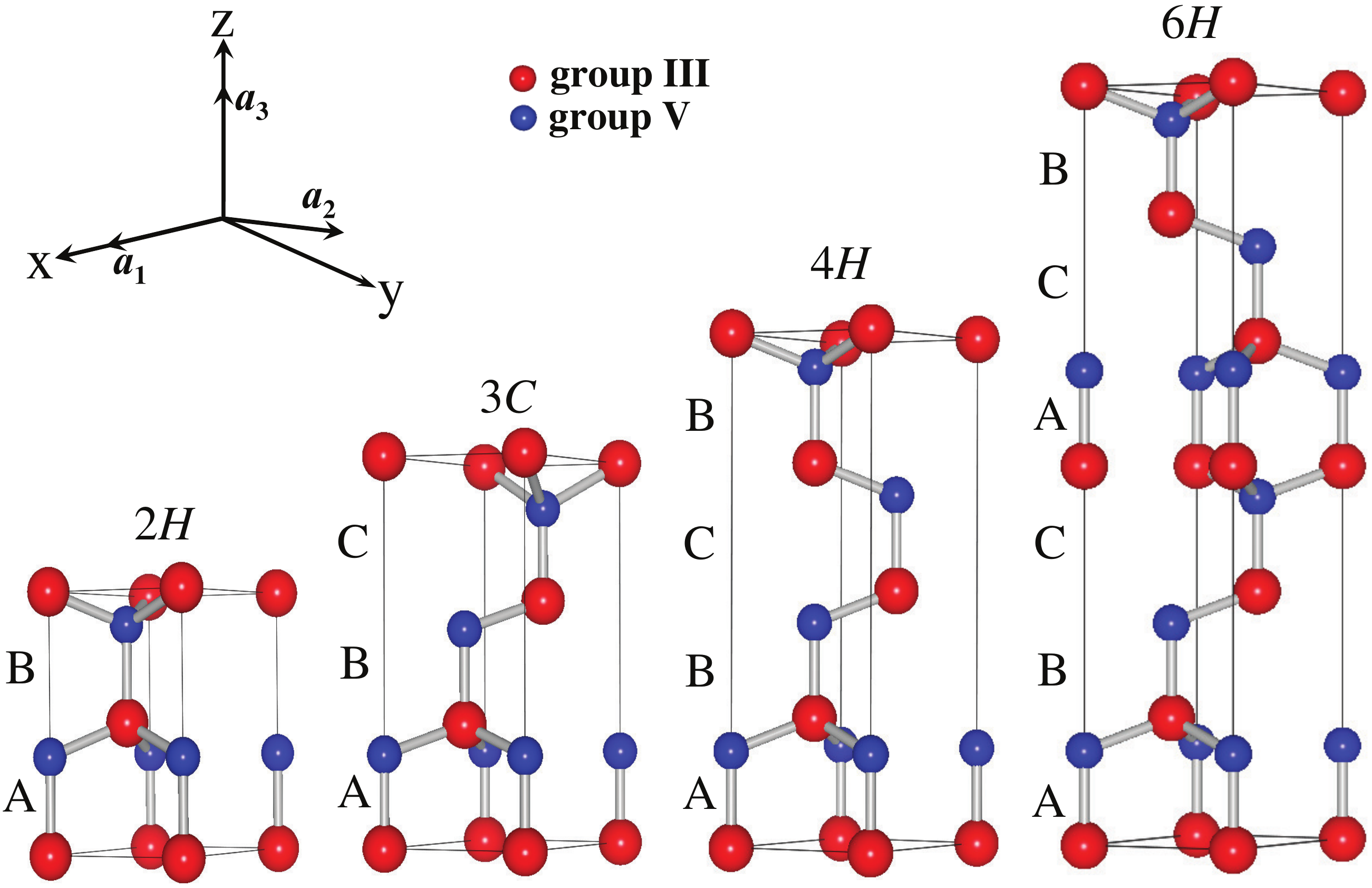}
\caption{(Color online) Stick-and-ball models of 3C and $p$H ($p=2,4,6$) polytypes. Cations: red spheres, anions: blue spheres. The stacking sequence of the cation-anion bilayers are indicated by the symbols A, B or C. Primitive unit cells are shown for the $p$H polytypes, while a non-primitive hexagonal cell is depicted to illustrate the 3C symmetry. The primitive basis vectors ${\textbf a}_i$ ($i=1,2,3$) are also shown.
\label{fig1}}
\end{figure}

Relatively little or almost nothing is known about the atomic geometries of hexagonal $p$H polytypes of non-nitride III-V compounds (see Fig.~\ref{fig1}). Only very recently lattice constants of 2H- and 4H-InAs and -InSb as well as 2H-GaAs and -InP have been published \cite{Kriegner.Panse.ea:2011:NL,McMahon.Nelmes:2005:PRL,Kriegner.ea:2012:nano}. Internal cell parameters are only measured for metastable bulk 2H-GaAs ($u=0.3693$) \cite{McMahon.Nelmes:2005:PRL} and pure 2H-InAs nanowires ($u=0.37502$) \cite{Zanolli.Pistol.ea:2007:JoPCM}. Mostly theoretical values are available for the 2H, 4H, and 6H polytypes of GaAs, InP, InAs, and InSb \cite{Panse.Kriegner.ea:2011:PRB}. We follow this line of structure calculations \cite{Panse.Kriegner.ea:2011:PRB} also for GaP and GaSb. 

The parameter-free total-energy and force calculations are performed in the framework of the DFT \cite{Hohenberg.Kohn:1964} within the local density approximation (LDA) \cite{Kohn.Sham:1965} as implemented in the Vienna \emph{ab initio} simulation package (VASP) \cite{Kresse.Furthmuller:1996}. The exchange-correlation (XC) functional is used as parameterized by Perdew and Zunger \cite{Perdew.Zunger:1981}. We do not take into account gradients of the electron density within the generalized gradient approximation (GGA), since LDA gives better structural parameters for conventional III-V compounds \cite{Haas.Tran.ea:2009:PRB}. The outermost $s$, $p$, and (in the case of Ga and In) $d$ electrons are treated as valence electrons whose interactions with the remaining ions is modeled by pseudopotentials generated within the projector-augmented wave (PAW) method \cite{Kresse.Joubert:1999}. The electronic wave functions between the cores are expanded in a basis set of plane waves. Its energy cutoff is tested to be sufficient with 500 eV for the six III-V compounds GaP, GaAs, GaSb, InP, InAs, and InSb under consideration. The Brillouin-zone (BZ) integrations are carried out on $\Gamma$-centered 10$\times$10$\times$$M$ {\bf k}-point meshes according to Monkhorst and Pack \cite{Monkhorst.Pack:1976} to achieve an overall energy convergence beneath 1 meV. The value of $M$ has to be varied according to the number of layers in stacking direction of the III-V polytype. We use $M=10,6,3,2$ for the 3C, 2H, 4H, and 6H polytype, respectively.

It is known but also confirmed by the computations \cite{Panse.Kriegner.ea:2011:PRB} that the DFT-LDA procedure gives rise to a minor underestimation of the lattice constants, e.g. for the cubic polytype of 0.8 \% (GaAs), 0.7 \% (InP), 0.4 \% (InAs), and 0.4 \% (InSb), in comparison to experimental (room-temperature) values \cite{Martienssen.Warlimont:2005:Book}. This underestimate may induce a small overestimation of the gaps of about $70-150$ meV taking the volume deformation potentials \cite{Yu.Cardona:1999:Book} into account. However, this deviation should not play a role for the band alignment, since similar variations of the lattice parameters are expected also for the hexagonal polytypes. Therefore, an error compensation is expected.

\begin{figure}[h!]
\hspace {-0.7 cm}
\includegraphics[width=0.35\textwidth, angle=-90]{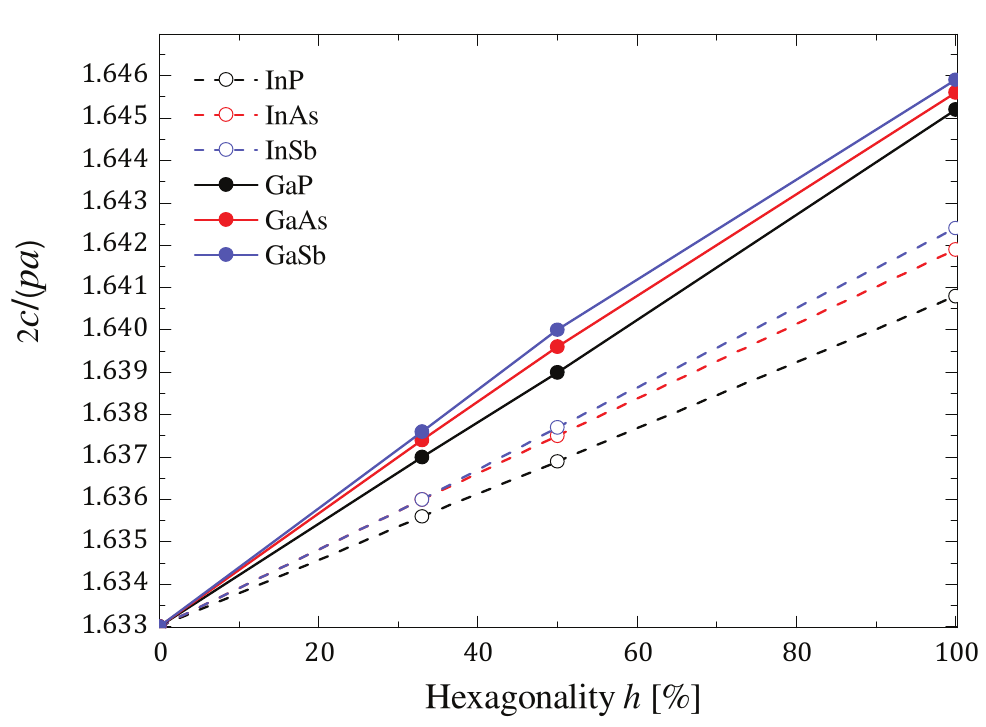}
\caption{Renormalized lattice constant ratio 2$c/(pa)$ versus polytype hexagonality $h$.
\label{fig2}}
\end{figure}

The structural parameters as the lattice constants $c$, $a$ and the resulting volume $V_{\rm pair}$ per cation-anion pair are given in Table I of supplementary information for the 3C and $p$H ($p=2,4,6$) polytypes of GaP and GaSb. Also the energy excess $\Delta E$ per pair with respect to the zinc-blende structure and the isothermal bulk modulus $B_{\rm 0}$ are listed. Together with the value of GaAs \cite{Panse.Kriegner.ea:2011:PRB} similar trends with the anion and the hexagonality as for the In-V compounds are observed for all structural $(a, c, V_{\rm pair})$, energetic  ($\Delta E$) and elastic ($B_{\rm 0}$) properties. This fact is clearly demonstrated in Fig.~\ref{fig2} for the $c/a$ ratio taking the results of Ref. [\onlinecite{Panse.Kriegner.ea:2011:PRB}] into account. Interestingly, the deviations from the ideal value 2$c/pa= \sqrt {8/3}$ are larger for Ga-V compounds in comparison to the In-V ones. This fact suggests that the hexagonal crystal field is larger in the case of the Ga cation compared to the In cation. 


\subsection{Quasiparticle bands}

Instead of the Kohn-Sham equation of DFT \cite{Kohn.Sham:1965} one has to solve a quasiparticle (QP) equation \cite{Aulbur.Jonsson.ea:1999:SSP} with a spatially non-local, non-Hermitian, and energy-dependent XC self-energy operator, e.g. within Hedin's GW approximation \cite{Hedin.Lundqvist:1970:InColl}. An efficient method to solve the QP equation has been recently developed \cite{Fuchs.Furthmuller.ea:2007:PRB,Bechstedt.Fuchs.ea:2009:PSSB}. Its iteration begins with a replacement of the XC self-energy by the functional derivative of the non-local HSE06 hybrid functional \cite{Heyd.Scuseria.ea:2006:JoCP,Krukau.Vydrov.ea:2006:JCP,Paier.Marsman.ea:2006} (using a parameter of $\omega=0.15$ a.u.$^{-1}$ instead of $\omega=0.11$ a.u.$^{-1}$, see disambiguation in Ref. \onlinecite{Paier.Marsman.ea:2006:a}). The next iteration step includes the deviation to the GW self-energy in a perturbation-theory manner. Therefore the method is called HSE06 + G$_0$W$_0$. In general, the described HSE06 + G$_0$W$_0$ QP method allows the parameter-free prediction of band structures for In- and Ga-V compounds with a high accuracy. \cite{Kim.Hummer.ea:2009:PRB,Kim.Marsman.ea:2010:PRB} When spin-orbit interaction \cite{Hobbs.Kresse.ea:2000:PRB} is included accurate band gaps of 1.48, 0.42, and 0.28 eV are obtained for InP, InAs, and InSb, respectively. \cite{Kim.Hummer.ea:2009:PRB}  Based on another hybrid functional PBE0 the corresponding QP computations yield 1.51 eV and 0.85 for GaAs and GaSb, repectively.\cite{Kim.Marsman.ea:2010:PRB} 

Unfortunately, hybrid-functional-based QP computations such as the HSE06 + G$_0$W$_0$ QP method are rather computer-time consuming. This holds especially for the 6H (4H) polytype whose unit cell contains 12 (8) atoms (see Fig.~\ref{fig1}). We have performed such computations (without spin-orbit interaction) only for the 3C and 2H polytypes of the six Ga and In phosphides, arsenides, and antimonides under consideration to have benchmark band structures for comparison. In order to perform converged QP calculations for all polytypes and compounds under consideration, we apply a recently developed slightly approximate QP method, the LDA-1/2 method \cite{Ferreira.Marques.ea:2008:PRB,Ferreira.Marques.ea:2011:AA}. This method allows the inclusion of spin-orbit interaction in a rather easy manner. In addition, it competes well with results of the GW QP approach. Fortunately, the computational effort of the LDA-1/2 method is the same as for the DFT-LDA method used to treat the ground-state properties.

The method is based on the previously successful density-functional technique of half-occupation \cite{Slater:1972:AQC}. Its principal idea goes back to Slater's transition state \cite{Slater.Johnson:1972:PRB,Leite.Ferreira:1971:PRA}. We apply this method by preparing a $pd$-like excitation in the electronic system of a certain compound for which XC is treated by the LDA functional \cite{Perdew.Zunger:1981}. In order to find a reasonable characterization of the excitation and a corresponding self-energy one needs occupation numbers and cutoff radii CUT for each atom. We construct them following the rules of (i) maximizing the fundamental gap of zinc-blende compounds,  (ii) transferability of the "atomic" parameters in different chemical environments, and (iii) the sum of radii CUT should be smaller than or of the same magnitude as a bond length. The values are listed in Table~\ref{tab2}. Only for the antimonides we have slightly changed the occupation from 50:50 to 70:30 for the cation:anion ratio because of their large spin-orbit effects.

\begin{table}
\begin{ruledtabular}
\caption{\label{tab2}
CUT parameters (in atomic units)and half-ionized orbitals used within the LDA-1/2 QP calculations .
}
\begin{tabular}{ccc} 
Atom & CUT (a.u.) & Half-ionized orbital \\ \hline
Ga & 1.23     & $d$ \\
In & 2.126 & $d$ \\
P  & 3.85       & $p$ \\
As & 3.86        & $p$ \\
Sb & 4.22        & $p$ \\ 
\end{tabular}
\end{ruledtabular}
\end{table}

The quality of the chosen parameters with respect to the fundamental gap of zinc-blende crystals is illustrated in Table~\ref{tab3}. The excellent agreement with (low-temperature) experimental values $E_g=2.35$ (GaP, indirect gap), 1.519 (GaAs), 0.812 (GaSb), 1.424 (InP), 0.417 (InAs), and 0.235 (InSb) eV \cite{Vurgaftman.Meyer.ea:2001:JAP} is obvious (see also supplementary information, Table II). The mean absolute relative error of the computed gaps amount to 2.9 \% and, hence, indicate a high predictive power of the LDA-1/2 method for the band structures of the polytypes around their fundamental gaps. In addition, we have also studied the effective electron and hole masses of zinc-blende III-V compounds near to $\Gamma$ (Table III of supplementary information). In general, we found exellent agreement for the electron, heavy-hole, and light-hole masses. Only the hole masses for the split-off band are slightly understimated with respect to experimental values. The mean deviation from the experimental mass values is equal or smaller the those derived from more sophisticated methods \cite{Kim.Hummer.ea:2009:PRB,Kim.Marsman.ea:2010:PRB}. For that reason we use the parameters in Table~\ref{tab2} also to predict the band structures and band parameters for the hexagonal polytypes $p$H ($p=2,4,6$).

\begin{table*}
\begin{ruledtabular}
\caption{\label{tab3}
Characteristic parameters (in eV) of the band structures and their alignments from LDA-1/2 QP calculations including spin-orbit interaction for four polytypes of six III-V compounds. The branch-point energyies $E_{BP}$ are given with respect to the valence band maximum. The positions of the band edges $E_c$ and $E_v$ use $E_{BP}$ as energy zero. The band offsets $\Delta E_c$ and $\Delta E_v$ are measured with respect to the band-edge position in the cubic 3C phase, $\Delta E_\nu=E_\nu(p$H$)-E_\nu$(3C) ($\nu=c,v$).
}
\begin{tabular}{c c c c c c c c c c c c c}

  & Compound& Polytype  & ${E_{g}}$ & ${\bigtriangleup_{cf}}$ & ${\bigtriangleup_{so}}$ & ${E_{BP}}$ & ${E_{v}}$ & ${E_{c}}$ & ${\bigtriangleup E_{v}}$ & ${\bigtriangleup E_{c}}$ & \\
  &&      $$   &   [eV]               &[eV]&    [eV]  &     [eV]    &[eV]& $[eV]$&[meV]&[meV]&\\ \hline
\\

&\textbf{GaP}&        3C                    & $(\Gamma$-$\Gamma)$      2.790       &   0.000      &  0.082     &0.735& -0.735 &2.055   & 0 & 0&
 \\ 
  &&&$(\Gamma$-$X)$ 2.330&&&&&1.595&0&-460&
\\  
&  &        6H               &$(\Gamma$-$\Gamma)$        2.322      &   0.021  &  0.083      & 0.690&-0.690&  1.632 & 45&-423 &
\\
&&&$(\Gamma$-$M)$  2.262&&&&&1.572&45&-483&
\\ 
 & &      4H                 &  $(\Gamma$-$\Gamma)$    2.267        &             0.027                 &              0.084    & 0.679&  -0.679&1.588   & 56& -467 &                                                                                                          
\\
&&&$(\Gamma$-$M)$ 2.270&&&&&1.591&56&-464&
\\ 
&  &        2H            &   $(\Gamma$-$\Gamma)$   2.181        &               0.045                &              0.083     & 0.600& -0.600 &1.581 & 135&-474 &
\\
&&&$(\Gamma$-$M)$2.266 &&&&&1.666&135&-389&
\\

\hline 
\\
&\textbf{GaAs}&        3C                &      1.421      &               0.000       &  0.348        & 0.541 & -0.541 &0.880  & 0 & 0&
 \\  
&  &        6H                &      1.439      &                0.052            &               0.348      & 0.503& -0.503   & 0.936  & 38 &  56 &
\\  
&  &      4H                  &      1.443        &               0.071           &                 0.348     & 0.486&-0.486    & 0.957 &55&  77 &                                                                                                          
\\
&  &        2H              &      1.453        &               0.129           &                 0.348     & 0.424& -0.424   &1.029 & 117& 149 &\\
\\
\hline 
\\

&\textbf{GaSb}&        3C                &     0.783      &               0.000       &  0.766        & 0.165 &-0.165  & 0.618 & 0 & 0&
 \\  
&  &        6H               &      0.806     &                0.057          &               0.770      & 0.138& -0.138  & 0.668  & 57&  50 &
\\  
&  &      4H                  &      0.814        &               0.076           &            0.771          & 0.127&-0.127  & 0.687 & 67 &  70 &                                                                                                          
\\
&  &        2H             &      0.835       &               0.148           &                 0.775    & 0.071& -0.071   &0.764 & 95& 146 &\\
\\
\hline 
\\

 &\textbf{InP} &        3C                 &      1.475       &               0.000             &                0.100         & 0.824 & -0.824&   0.651& 0 & 0&
 \\  
  &&        6H                &      1.518      &                0.029            &                0.099      & 0.817& -0.817 &  0.701& 7& 50 &
\\  
  &&      4H                  &      1.533        &               0.039           &                 0.100     & 0.805 &  -0.805& 0.728  & 19&  77 &                                                                                                          
\\
 & &        2H            &      1.576        &               0.062           &                 0.104     & 0.742 & -0.742& 0.834 & 82& 183 &\\
\\ \hline  
\\
&\textbf{InAs} &        3C                &      0.411       &               0.000             & 0.357   & 0.596 &-0.596 &-0.185&  0 & 0&
 \\  
&  &        6H                &      0.431      &                0.034            &                0.350      & 0.582&-0.582 & -0.151& 14& 34&
\\  
&  &      4H                  &      0.440        &               0.055           &                0.356     & 0.567& -0.567 &-0.127& 29& 58 &                                                                                                          
\\
 & &        2H              &     0.481        &               0.095           &                0.356     & 0.540& -0.540&-0.059 & 56& 126&\\
\\
\hline 
\\
&\textbf{InSb}&        3C                &      0.230      &               0.000             &                0.772       & 0.286& -0.286& -0.056& 0 & 0&
 \\  
 & &        6H                &      0.244      &                0.036           &                0.772      & 0.261& -0.261& -0.017 & 25&  39&
\\  
  &&      4H                &      0.249        &               0.061          &                 0.772    & 0.246& -0.246&  0.003 & 40& 59 &                                                                                                          
\\
 & &        2H              &      0.264       &               0.113           &                 0.771     & 0.207& -0.207& 0.057& 79& 113 &\\

\end{tabular}
\end{ruledtabular}
\end{table*}

\subsection{Band alignment}

In order to determine the band discontinuities $\Delta E_c$ and $\Delta E_v$ for a heteromaterial or here heterocrystalline junction, one needs an alignment of the energy scales and hence the band structures on both sides of such a junction. The computational method for a more or less lattice-matched heterojunction uses the electrostatic potentials across the junction and those of the two materials \cite{VandeWalle.Martin:1986:PRB,Lambrecht.Segall.ea:1990:PRB}. This procedure however requires the construction of a certain interface between two polytypes. We neglect the small effects due to the interface, e.g. the interface dipole. We apply a more "macroscopic" approach \cite{Schleife.Fuchs.ea:2009:APL}, which only requires the calculation of the QP band structures of the adjacent polytypes. It asks for a universal reference level. Frensley and Kroemer \cite{Frensley.Kroemer:1976} suggested the use of an internal reference level, e.g. the branch-point energy $E_{\rm BP}$ \cite{Schleife.Fuchs.ea:2009:APL,Tersoff:1984:PRB}. This is the energy at which the band states change their character from predominantly acceptor-like (usually valence-band) states to mostly donor-like (usually conduction-band) states. We calculate $E_{\rm BP}$ by means of an approximate method \cite{Schleife.Fuchs.ea:2009:APL} that was successful for several material combinations \cite{Schleife.Fuchs.ea:2009:APL,Hoffling.Schleife.ea:2010:APL} and heterocrystalline systems \cite{Belabbes.Carvalho.ea:2011:PRB}.

\section{Band structure of polytypes}
\subsection{Bands: An overview}

\begin{figure*}
\includegraphics[width=1.25\textwidth, angle=-90]{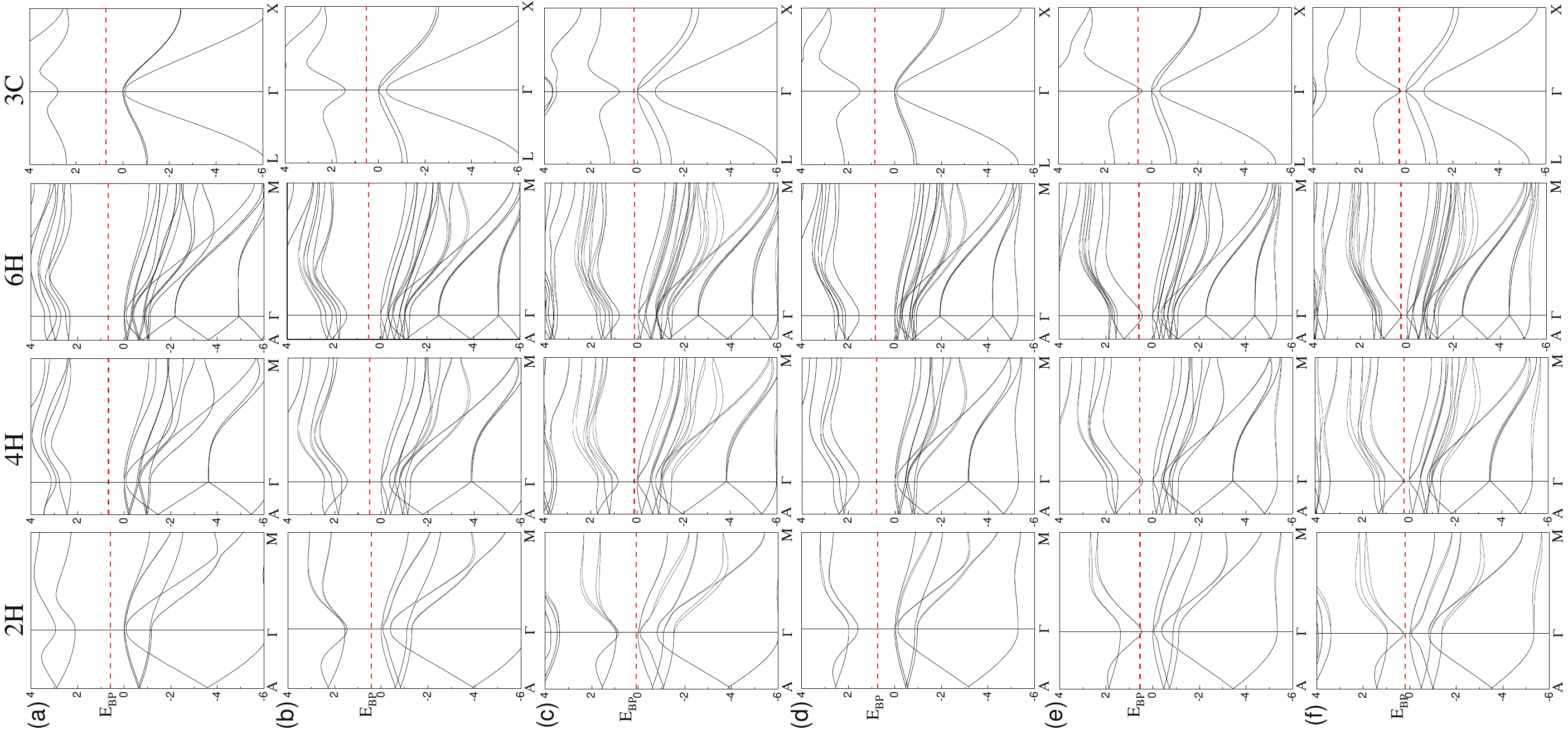}
\caption{Quasiparticle band structures including spin-orbit interaction $vs$ two high-symmetry lines in the cubic or hexagonal BZ. (a) GaP, (b) GaAs, (c) GaSb, (d) InP, (e) InAs, and (f) InSb. The four panels depict the bands for the 3C, 6H, 4H, and 2H polytypes. The valence band maximum is used as energy zero. The branch-point energy is indicated by a horizontal red dashed line.
\label{fig3}}
\end{figure*}

\begin{figure*}
\hspace{-0.5cm}
\resizebox{2.1\columnwidth}{!}{
\includegraphics[angle=-90]{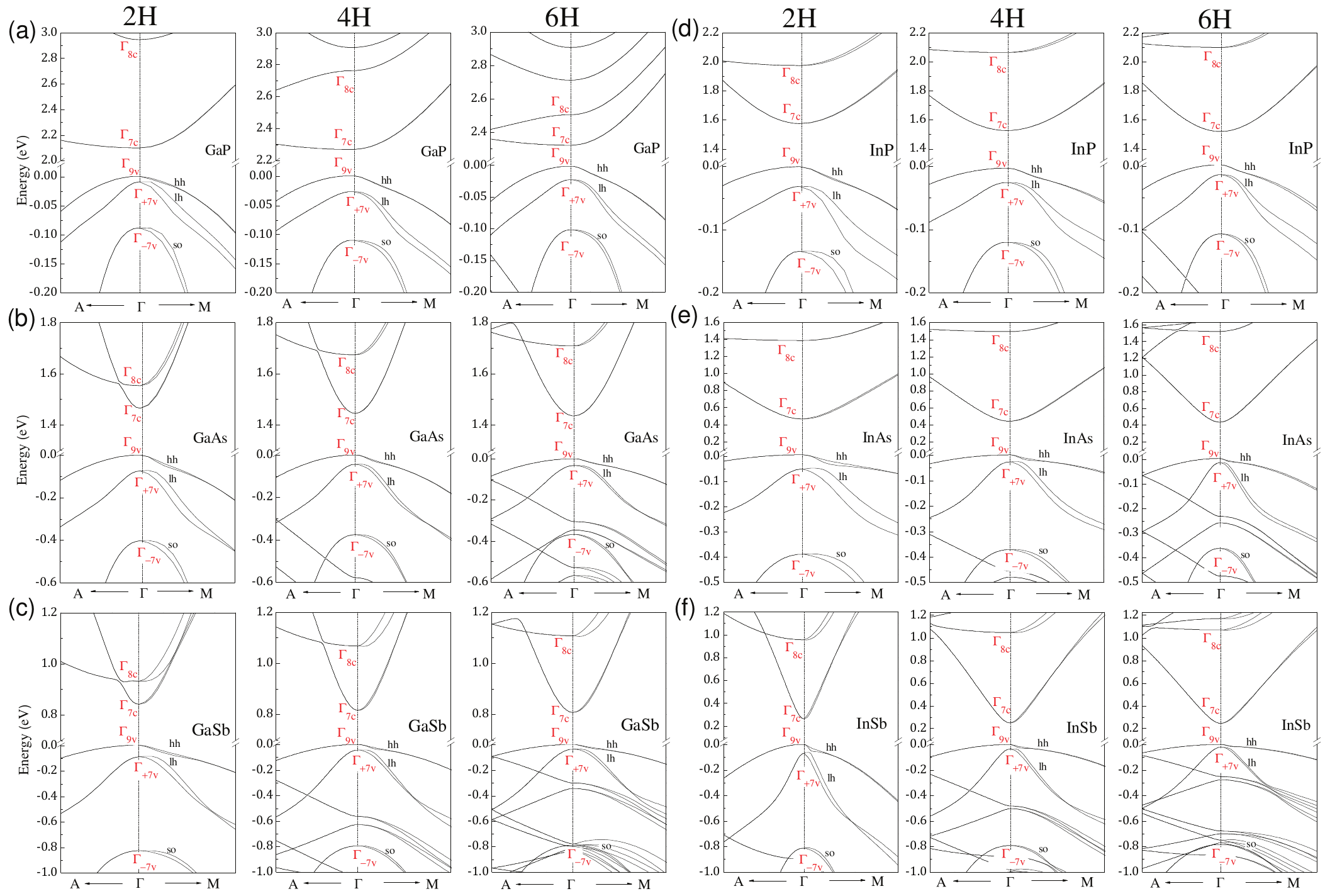}
}
\caption{As Fig.~\ref{fig3} but only the uppermost valence and lowest conduction bands near $\Gamma$ of the hexagonal polytypes  (a) GaP, (b) GaAs, (c) GaSb, (d) InP, (e) InAs, and (f) InSb. The valence band maximum is used as energy zero. The symmetry of the most important states is indicated.
\label{fig4}}
\end{figure*}

The QP band structures including spin-orbit interaction of the six III-V compounds under consideration as obtained within the above-described LDA-1/2 method are plotted in Fig.~\ref{fig3}.
Details of the uppermost valence and lowest conduction bands near the BZ center $\Gamma$ are shown in Fig.~\ref{fig4}.

First of all, the QP band structures in Fig.~\ref{fig3} for the zinc-blende polytype are in good agreement with previous QP results for In compounds \cite{Kim.Hummer.ea:2009:PRB} and known general behavior of the band ordering (see e.g. \cite{Yu.Cardona:1999:Book,Kim.Marsman.ea:2010:PRB}), and the experimental values for the most important bands at high-symmetry points $\Gamma$,$X$, and $L$ (see collection in Ref. [\onlinecite{Kim.Marsman.ea:2010:PRB}]). This especially holds for the position of the band extrema at the $X$ and $L$ points. Most important we clearly confirm that apart from GaP the $L_{6c}$ level is below the $X_{6c}$ one. This ordering has consequences for the interpretation of the polytype bands. The valence band maximum (VBM) $E_v$ in zinc blende is of $\Gamma_{8v}$ type. Apart from GaP where the conduction band minimum (CBM) $E_c$ is situated at the $X$ point, for all other compounds the CBM possesses $\Gamma_{6c}$ symmetry indicating their direct character. Only 3C-GaP represents an indirect semiconductor with the indirect gap $E_g$ ($X_{6c}-\Gamma_{8v}$) and the direct gap $E_g$ ($\Gamma_{6c}-\Gamma_{8v}$). The spin-orbit-interaction-induced splittings of degenerate valence band states but also the wave-vector-induced band splittings along the $\Gamma L$ direction increase with the size of the anion while the influence of the Ga or In cation is depressed.

Principal features of the band structures of the $p$H polytypes in Fig.~\ref{fig3} can be understood by simple folding arguments. For instance, to understand the lowest conduction bands one has to fold the $L_{6v}$  zinc-blende state onto the $\Gamma$ point, giving rise to the $\Gamma_{8c}$ state in wurtzite crystals. It is usually above the pure $s$-like state $\Gamma_{7c}$ (which arises from the $\Gamma_{6c}$ in the zinc-blende case). However, due to the slightly changed bonding behavior in the hexagonal 2H crystal the energetical order of the two levels $\Gamma_{1c}$ and $\Gamma_{3c}$ without spin-orbit interaction ($\Gamma_{7c}$ and  $\Gamma_{8c}$ with spin-orbit inetraction) depends sensitively on the atomic geometry and the strain state as recently demonstrated for GaAs \cite{Cheiwchanchamnangij.Lambrecht:2011:PRB}. Also in Figs.~\ref{fig3} and ~\ref{fig4} the two conduction bands are close to each other for 2H-GaAs. In any case, we state a clear contradiction to the empirical pseudopotential results \cite{De.Pryor:2010:PRB}. De and Pryor claim that in all 2H-Ga compounds a band inversion occurs so that the $\Gamma_{8c}$ level is below the $\Gamma_{7c}$ one. This result is obviously a consequence of the chosen symmetric and antisymmetric psuedopotential form factors and the wrong crystal-field splittings.

In general, the situation of the $\Gamma_{7c}/\Gamma_{8c}$ band ordering and hence the band distance $\Delta_{CB}=E_c(\Gamma_{7c})-E_c(\Gamma_{8c})$ in GaAs are under debate. Theoretical values amount to $\Delta_{CB}=-23$ meV \cite{Murayama.Nakayama:1994:PRB}, $\Delta_{CB}=+85$ meV \cite{De.Pryor:2010:PRB}, and $\Delta_{CB}=-81$ meV \cite{Cheiwchanchamnangij.Lambrecht:2011:PRB}, while our value is $\Delta_{CB}=-85$ meV. Recent resonant Raman spectroscopy experiments of 2H-GaAs clearly showed that the conduction band minimum is of $\Gamma_{7c}$ symmetry \cite{Ketterer:NANO}  in agreement with our prediction but in disagreement with empirical pseudopotential studies.\cite{De.Pryor:2010:PRB}
In the case of 4H (6H) the lowest conduction band of 3C at $L$ and 0.5 $\Gamma L$ (2/3 $\Gamma L$ and 1/3 $\Gamma L$) is folded onto the $\Gamma$ point. As a result besides the $\Gamma_{7c}$ level $(p-1)$ (twofold-degenerate) conduction levels appear nearby in the case of a $p$H polytype. This can be clearly seen in Fig.~\ref{fig3} e.g. for InAs and InSb. Of course, the lowest conduction bands of GaP polytypes are more difficult to explain using simple folding arguments due to the reverse ordering of the conduction-band minima.

At first glance the uppermost valence bands at $\Gamma$ of the hexagonal crystals are similar to that of 3C. Only the (positive) crystal-field splitting $\Delta_{cf}$ (see Table~\ref{tab3}) leads to an additional splitting of the $\Gamma_{8v}$ state in 3C besides the $\Gamma_{8v}-\Gamma_{6v}$ splitting due to the spin-orbit interaction. In wurtzite crystals one expects a sequence of the valence levels $\Gamma_{9v}$, $\Gamma_{7v+}$, and $\Gamma_{7v-}$, which is present in Figs.~\ref{fig3} and \ref{fig4} in agreement with the empirical-pseudopotential results \cite{De.Pryor:2010:PRB}. For the 4H and 6H polytypes of the arsenides and antimonides a problem arises due to the two (with spin four) relatively flat valence bands along the $\Gamma L$ line in 3C. As shown in Fig.~\ref{fig4} the uppermost twofold (with spin) degenerate levels $\Gamma_{9v}$ and $\Gamma_{7v+}$ can still be clearly identified. However, while in the 2H case the $L_{4,5v}$ and $L_{6v}$ levels are folded onto $\Gamma_{8v}$ and $\Gamma_{9v}$ states below $\Gamma_{7v-}$, the valence band states from $\frac{1}{2}\Gamma L$ (4H) or $\frac{1}{3}\Gamma L$ and $\frac{2}{3}\Gamma L$ (6H) are folded onto energies at the $\Gamma$ point near to the $\Gamma_{7v-}$ level (4H) or even above it. Therefore, we did a careful symmetry analysis of the valence states at $\Gamma$ to identify the $\Gamma_{7v-}$ band which mainly consists of atomic $p_z$-like orbitals. The figure panels for antimonides show that the fifth (seventh) twofold degenerate level below VBM corresponds to $\Gamma_{7v-}$ in the 4H (6H) case.

\subsection{Fundamental gaps}

\begin{figure}[h!]
\includegraphics[width=0.35\textwidth, angle=-90]{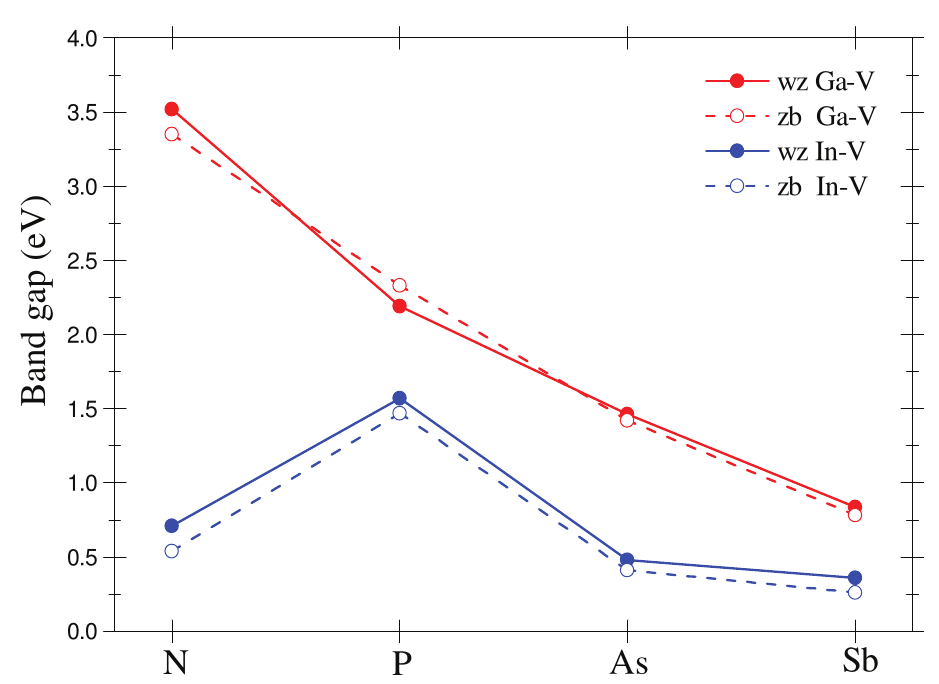}
\caption{The gaps $E_g({2H}) $ and  $E_g(\text{3C})$ versus the group-V anion are indicated by solid or dashed lines.
\label{fig5}}
\end{figure}

The Figs.~\ref{fig3} and \ref{fig4} and Table~\ref{tab3} indicate a clear trend of the fundamental energy gaps $E_g$ with the hexagonality $h=$  0\% (3C), 33\% (6H), 50\% (4H), and 100\% (2H). Apart from 6H-GaP all hexagonal polytypes represent direct semiconductors. The $E_g$ values monotonously increase with rising $h$. There is only one discrepancy from this trend when going from 3C- to 2H-GaP. The reason is related to the folding of the uppermost conduction band in the 3C polytype along $\Gamma L$ onto the $\Gamma$ point of the hexagonal BZs and the indirect character of 3C-GaP. 
The frequently asked question how the band gaps in 3C and 2H relate to each other is clearly answered by Table~\ref{tab3}. Apart from the indirect semiconductor GaP the band gaps in wurtzite are larger than those in zinc blende. We observe a clear trend of the absolute variations of the gaps going from 3C to 2H: -149 (GaP), 32 (GaAs), 52 meV (GaSb), 101 (InP), 70 (InAs), and 34 meV (InSb).
The polytypic gap splittings for GaN and InN \cite{Carvalho.Schleife.ea:submitted:PRB,Belabbes.Carvalho.ea:2011:PRB} also support this trend (see Fig.~\ref{fig5}). Figure~\ref{fig5} confirms the gap increase with the hexagonality for In-V and Ga-V compounds. Only GaP shows an opposite behavior.
Our results are in agreement with the usual trend in other compounds such as nitrides and oxides where the $zb$-$wz$ polytypism has been observed. A rough rule can be derived that the gap difference decreases with increasing size of the anion where as the opposite trend is valid for the cations. The anomalous trend of the gaps versus the anion for InN has been explained elsewhere. \cite {carrier:033707}

We state that at least the increase of the gap with the hexagonality is in qualitative agreement with experimental results. Photoluminescence (PL) measurements of Spirkoska et al. \cite{Spirkoska.Arbiol.ea:2009:PRB} and Hoang et al. \cite{Hoang.Moses.ea:2009:APL} indicate an increase of the gap from 3C- to 2H-GaAs of 33 meV in excellent agreement with our predictions. Photoluminescence \cite{Spirkoska.Arbiol.ea:2009:PRB}, Photoluminescence excitation spectroscopy \cite{Ketterer.Heiss.ea:2011:PRB}  and resonant  Raman spectroscopy \cite{Ketterer:NANO} support the increase of the 2H-GaAs gap with respect to its 3C value. Such PL measurements performed by almost the same group however indicate an opposite shift of $- 23$ meV \cite{Heiss.Conesa-Boj.ea:2011:PRB}. Very recent  luminescence studies also support ${E_{g}}$(3C) $ <$  ${E_{g}}$(2H) for GaAs \cite{Spirkoska:2012:PRB,Jahn:2012:PRB}.
The PL studies of InP nanowires suggest an increase larger than 36 meV \cite{Akopian.Patriarche.ea:2010:NL} in qualitative agreement with our computations. Tight-binding calculations \cite{Akiyama.Yamashita.ea:2010:NL}
 qualitatively support our findings with a gap increase of 110 meV for InP.

\subsection{Valence-band parameters}

\begin{figure*}
\includegraphics[width=0.385\textwidth, angle=-90]{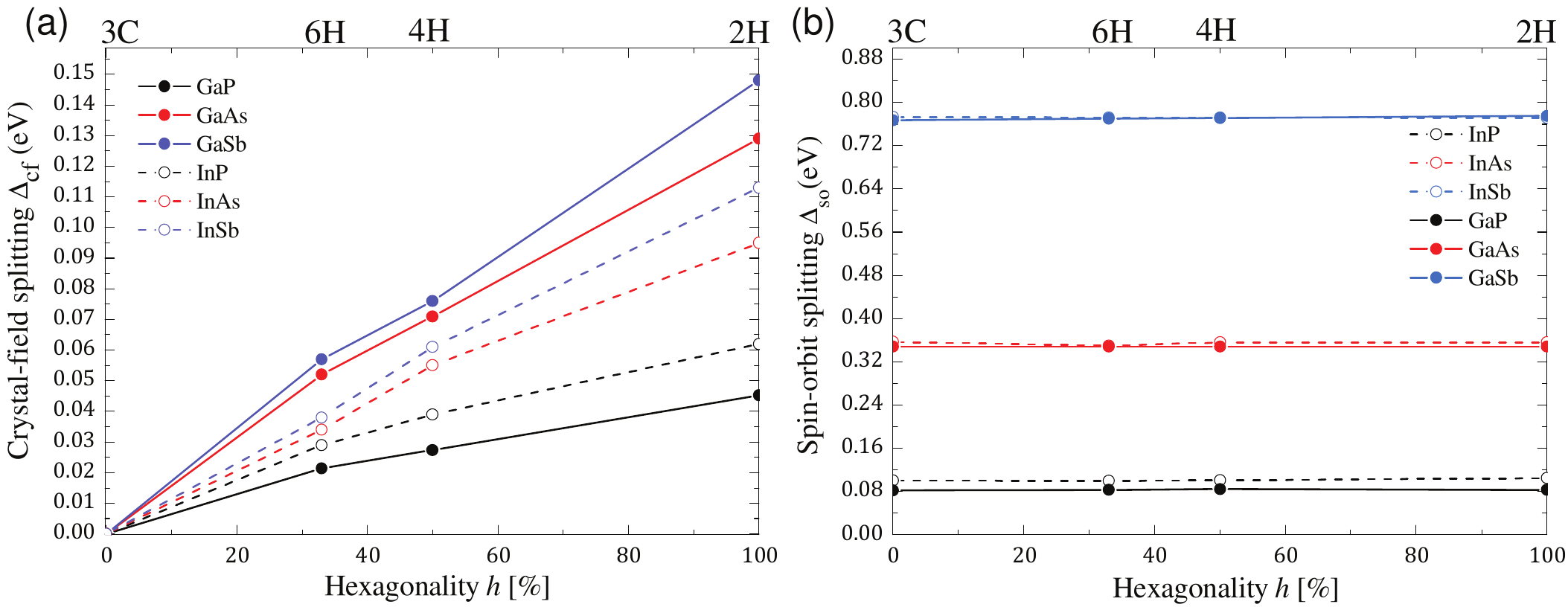}
\caption{Crystal-field (a) and spin-orbit (b) splitting versus the polytype hexagonality $h$.
\label{fig6}}
\end{figure*}

From the band structures in Figs.~\ref{fig3} and \ref{fig4} we derive the most important splitting parameters of the valence bands. Within the quasicubic approximation (where the anisotropy of the spin-orbit interaction in the $p$H polytypes is neglected) only the crystal-field splitting $\Delta_{cf}$ (characterizing the hexagonal crystal field) and the spin-orbit splitting $\Delta_{so}$ of the pure $p$-states are relevant. The ${\bf k}\cdot{\bf p}$ perturbation theory \cite{Chuang.Chang:1996:PRB} gives
\begin{align}\label{eq1}
\begin{split}
    E_v(\Gamma_{9v})-E_v(\Gamma_{7v\pm})= \\ \frac{1}{2}\bigg[\left(\Delta_{cf}+\Delta_{so}\right)\mp\sqrt{\left(\Delta_{cf}-\frac{1}{3}\Delta_{so}\right)^2+\frac{8}{9}\Delta^2_{so}}\bigg].
\end{split}
\end{align}
Solving the identification problem of the $\Gamma_{7v-}$ valence state discussed above in Fig.~\ref{fig4} , the parameters $\Delta_{cf}$ and $\Delta_{so}$ can be determined. The results are also listed in Table~\ref{tab3} and depicted in Fig.~\ref{fig6}.

For each compound the crystal-field splitting $\Delta_{cf}$ increases monotonously with the polytype hexagonality $h$ (see Fig.~\ref{fig6}a and Table~\ref{tab3}) as expected.
This is in agreement with the increase of the aspect ratio $c/a$ (see Fig.~\ref{fig2} and also \cite{Kriegner.Panse.ea:2011:NL,Panse.Kriegner.ea:2011:PRB}) and the deviation of $u$ (not explicitly given) from its ideal value $u=0.375$.
In contrast to the nitrides \cite{Carvalho.Schleife.ea:submitted:PRB} the $u$ parameters computed for the other Ga-V and In-V compounds fulfill the condition $u<0.375$. Consequently, a clear increase of $\Delta_{cf}$ is found along the anion row P, As, and Sb, i.e., with the anion size. The opposite fact holds for the cations, only GaP deviates.  Also the $\Delta_{cf}$ value for InN \cite{Carvalho.Schleife.ea:submitted:PRB} supports this trend along the group-V anions. The computed absolute value $\Delta_{cf}=129$ meV for 2H-GaAs is only slightly smaller than the value 180 meV derived within GW calculations. \cite{Zanolli.Fuchs.ea:2007} The chemical trends and especially the absolute magnitude of the $\Delta_{cf}$ values derived from empirical pseudopotential calculations \cite{De.Pryor:2010:PRB} are in disagreement with our findings. Crystal-field splittings for the phosphides much larger than 0.3 eV seem to be rather unrealistic in comparison to the well-accepted values for nitrides. \cite{Carvalho.Schleife.ea:submitted:PRB} Only the 2H-InSb value 159 meV \cite{De.Pryor:2010:PRB} approaches the splitting given in Table~\ref{tab3}.

Results of the fitting procedure with formula~\eqref{eq1} for the spin-orbit splitting (neglecting its anisotropy) are given in Table~\ref{tab3} and Fig.~\ref{fig6}b. Taking the accuracy of the computations into consideration as a general result we find that the spin-orbit splitting is rather independent of the polytype. As a long-range interaction the hexagonal crystal field does not influence the spin-orbit coupling constant for the valence $p$ electron states. Our values are in complete agreement with those of previous calculations \cite{De.Pryor:2010:PRB,Lambrecht.Segall.ea:1990:PRB}. We also state excellent agreement with spin-orbit splittings of $\Delta_{so}=80$ (GaP), 341 (GaAs), 760 (GaSb), 108 (InP), 390 (InAs), and 810 meV (InSb) measured for the zinc-blende polytype. \cite{Vurgaftman.Meyer.ea:2001:JAP} Spin-orbit splitting $\Delta_{so}=379$ meV and crystal-field splitting $\Delta_{cf}=189$ meV
derived from resonance Raman scattering mesurements of 2H-GaAs \cite{Ketterer:NANO} slightly overstimate the calculated values in Table~\ref{tab3}.
\section{Band offsets}
\subsection{Branch-point energies}

In order to align to the band structures depicted in Figs.~\ref{fig3} and \ref{fig4} for the different polytypes of the Ga and In compounds we use the branch-point energy $E_{BP}$. We apply a recently developed approximate method to compute these energies from the known QP band structures \cite{Schleife.Fuchs.ea:2009:APL}. It is slightly generalized in order to take all the spin-orbit split bands into account. Basically the number of conduction and valence bands used in the computation has been doubled. Results with respect to the $\Gamma_{8v}$ (3C) or $\Gamma_{9v}$ ($p$H) VBM are listed in Table~\ref{tab3} together with the band extrema $E_c$ and $E_v$ referred to $E_{BP}$ as energy zero.

There are clear chemical trends for the $E_{BP}$ values measured with respect to the VBM versus the anion and the hexagonality. This is also true for their variation with the crystal structure. In general, the variation of $E_{BP}$ is much smaller than the variation of the fundamental energy gaps $E_g$. It is restricted to the interval 0.1 eV$<E_{BP}<$0.8 eV. As a consequence $E_{BP}$ generally represents a midgap level for InP, 4H- and 2H-InSb, GaP, GaAs,  and GaSb. Independent of the polytype the level $E_{BP}$ appears deep in the conduction band for InAs and the low-hexagonality InSb polytypes. Such a behavior is well known for InN \cite{King.Veal.ea:2007:a,King.Veal.ea:2008}. As a consequence a surface $n$-accumulation layer (also on the nanowire surface) should occur. Such a surface accumulation layer has been experimentally observed not only for InN \cite{King.Veal.ea:2007:a,King.Veal.ea:2008} but also for InAs \cite{Noguchi.Hirakawa.ea:1991:PRL}. Our results for the branch point in zinc-blende polytypes are in qualitative and even quantitative agreement with those of other calculations for the charge-neutrality level for the six III-V compounds under consideration \cite{Tersoff:1984:PRB,Robertson.Falabretti:2006:JoAP,monch:2001:JAP}. For the Ga-V compounds the computed $E_{BP}$  values are also in excellent agreement with the charge-neutrality levels 1.00 eV (GaP), 0.54 eV (GaAs), and 0.07 eV (GaSb) derived from Schottky barrier height data (see collection in Ref. [\onlinecite{monch:2001:JAP}])
\
There is another indication for the reliability of the branch-point energies in Table~\ref{tab3} for alignment when comparing the III-V compounds in their zinc-blende geometry. The absolute position of the VBM increases for Ga-V and In-V with the anion from P to As and to Sb. This is in qualitative agreement with measured ionization energies \cite{monch:2001:Book} and measured band discontinuities with respect to Si and Ge \cite{Katnani:1983,Margaritondo:1985:PRB}.

\subsection{Band lineup}

\begin{figure*}
\includegraphics[width=1\textwidth, angle=-90]{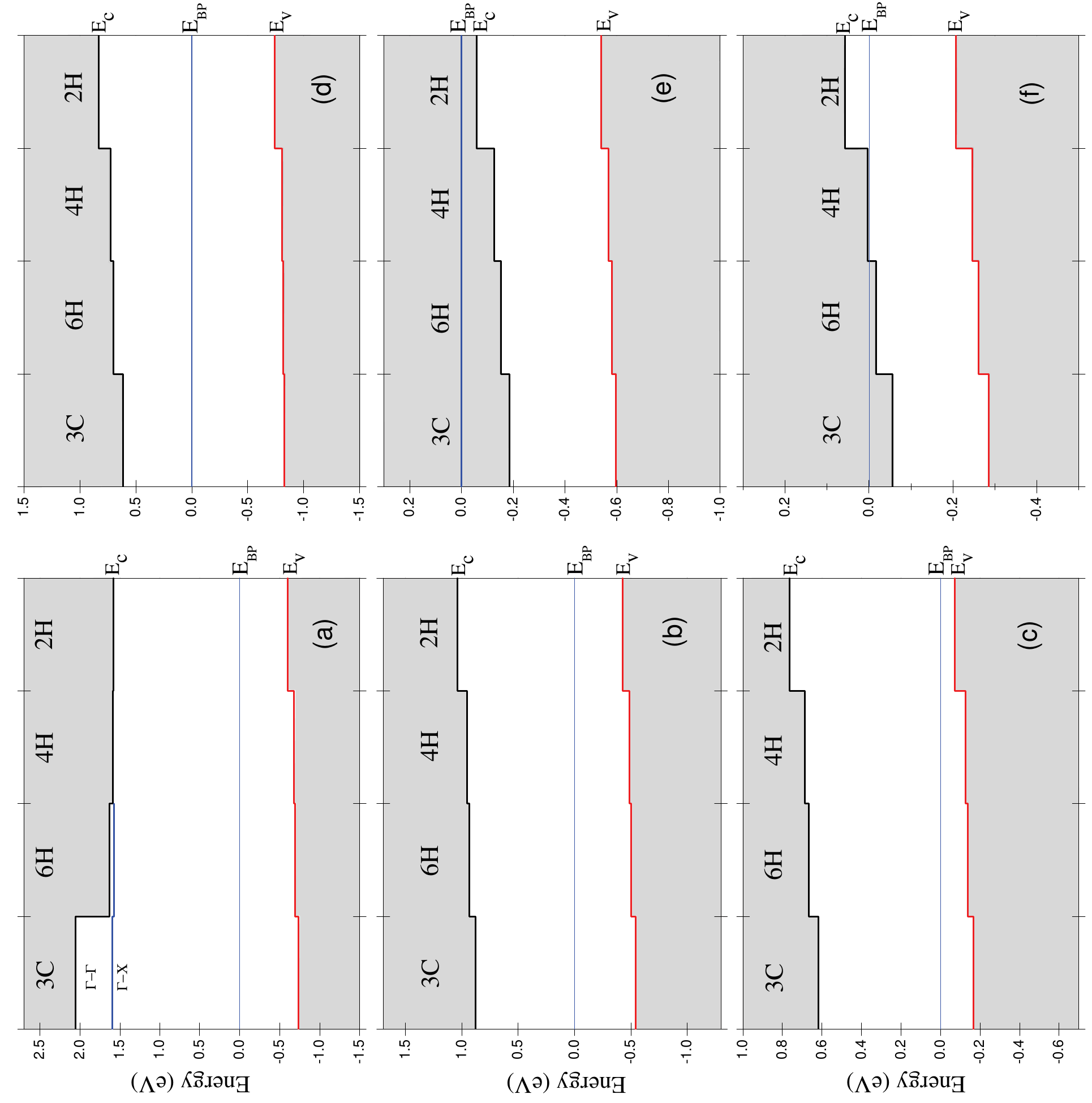}
\caption{Band lineups for the four polytypes 3C, 6H, 4H, and 2H with increasing hexagonality of III-V compounds. The conduction band minimum $E_c$ (black), the valence band maximum $E_v$ (red), and the branch-point energy (blue) are depicted for (a) GaP, (b) GaAs, (c) GaSb, (d) InP, (e) InAs, and (f) InSb. The shaded energy regions indicate the allowed bands.
\label{fig7}}
\end{figure*}
The branch-point alignment of the band edges of the four polytypes leads to the band edges $E_c$ and $E_v$ (with respect to the $E_{BP}$ energy zero) and, consequently, the band discontinuities $\Delta E_c$ and $\Delta E_v$ given all in Table~\ref{tab3}. They are used to plot the band lineups in Fig.~\ref{fig7}. Omitting for a moment GaP, where the discussion of the conduction band offsets is more difficult due to the indirect character of the 3C  and 6H polytypes, some general rules can be derived for the other III-V compounds. All the heterotransitions 3C-$p$H and $p'$H-$p$H with $p'>p$ ($p, p'$=2, 4, 6) represent type-II structures with a staggered arrangement of the band edges $E_c$ and $E_v$ \cite{Yu.Cardona:1999:Book,Kittel:2005:Book}. The band discontinuities $\Delta E_c$ and $\Delta E_v$ with respect to the cubic polytypes 3C rise monotonously with the hexagonality for each compound. This tendency is in line with the increase of the gaps with the hexagonality $h$ (see Table~\ref{tab3} and  Fig.~\ref{fig5}). However, because of the type-II character of the heterocrystalline structures the variation of  $E_c$ and $E_v$ is larger than that of  the gaps $\Delta E_g$  (with respect to 3C). For GaP the variation of $\Delta E_v$ is  similar while the position of the lowest conduction band minimum, independent of the directness or indirectness of the polytype, is rather constant with respect to the branch-point energies. The absolute values $\Delta E_c$  and $\Delta E_v$ decrease with the rising size of the anion as well as cation. The exception is  $\Delta E_v$ from InAs to InSb. This fact seems to be a consequence of the strong increase of the spin-orbit splitting constant  $\Delta_{so}$ for the valence bands. 

\subsection{Comparison with other calculations and measurements}

The staggered type-II character with the 3C valence band $E_v$ as the lowest occupied level is confirmed by a series of spectroscopic measurements for the 3C-2H (or $p$H, in general heterocrystalline junction with hexagonal stacking) of GaAs \cite{Spirkoska.Arbiol.ea:2009:PRB,Heiss.Conesa-Boj.ea:2011:PRB} and InP \cite{Akopian.Patriarche.ea:2010:NL}, and InAs \cite{Caroff.Dick.ea:2009:NN,Caroff.Bolinsson.ea:2011:STQE}. However, there are only a few values for a quantitative comparison. From PL measurements \cite{Heiss.Conesa-Boj.ea:2011:PRB} values $\Delta E_v=76\pm12$ meV and $\Delta E_c=53\pm20$ meV have been derived for the 3C-2H GaAs junction. The order of magnitude is in agreement with our predictions. The direct comparison of theory and experiment is however difficult because of several facts: (i) Theory only computes so-called 'natural' band discontinuities without taking into account the interface between bulk polytypes. (ii) The measurements are influenced by the real bilayer stacking in the studied nanowire and the confinement of electrons or holes. Table~\ref{tab3} makes obvious that a reduction of the hexagonality of the stacking sequence significantly reduces the band offsets. (iii) Moreover, type-II structures favor optical transitions which are indirect in space.

The comparison with other theoretical values is mainly restricted to the DFT-LDA method which suffers from the gap underestimate and takes no spin-orbit interaction into account. Heiss et al. \cite{Heiss.Conesa-Boj.ea:2011:PRB} however give values $\Delta E_v=122$ meV and $\Delta E_c=101$ meV for GaAs not too far from those in Table~\ref{tab3}. A more complete collection of values is given by Murayama and Nakayama \cite{Murayama.Nakayama:1994:PRB} for 3C-2H with $\Delta E_c=126$ (GaP), 117 (GaAs), 102 (GaSb), 129 (InP), 86 (InAs), and 86 meV (InSb) as well as $\Delta E_v=81$ (GaP), 84 (GaAs), 89 (GaSb), 45 (InP), 46 (InAs), and 57 meV (InSb), which  are significantly smaller than the values in the  Table~\ref{tab3}. However, in this paper the band alignment has been made by aligning the center of gravity for the uppermost three valence bands at $\Gamma$.

\section{Summary and conclusions}

Using the LDA-1/2 method, an approximative scheme to compute quasiparticle electronic stuctures, and taking the spin-orbit interaction into account, we have studied the quasiparticle band structures of the 6H, 4H, and 2H (wurtzite) polytypes of conventional III-V compounds which crystallize in the 3C (zinc blende) structure under ambient conditions. Using folding and symmetry arguments the valence band maxima have been found to be $\Gamma_{9v}$ for the hexagonal polytypes of all compounds studied. Apart from GaP the energetic ordering of  the lowest conduction bands is fixed to $\Gamma_{7c}$ below $\Gamma_{8c}$, although these levels are close for GaAs.

Band lineups for heterojunctions 3C-$p$H or $p'$H-$p$H ($p,p'=2,4,6$) have been predicted using the branch-point energy as common reference level. Apart from GaP, all other compounds give rise to staggered type-II junctions. Thereby the variation of the band edges is proportional to the hexagonality (i.e., the stacking) difference between the polytypes forming the junction. The comparison with recent measurements show qualitative and quantitative agreement.

\section*{Acknowledgements}

We acknowledge financial support from the Fonds zur F\"orderung der Wissenschaftlichen Forschung (Austria) in the framework of SFB 25 'Infrared Optical Nanostructures', and the EU ITN RAINBOW (Grant No. 2008-2133238).

\bibliography{literatur}

\end{document}